\documentclass[a4paper,fleqn,usenatbib]{mnras}

\usepackage{newtxtext,newtxmath}
\usepackage[T1]{fontenc}
\usepackage{ae,aecompl}

\usepackage{bm}
\usepackage{ulem}

\newcommand\plotone[1]{
 \centering
 \includegraphics[width=\columnwidth]{#1}
}

\newcommand\plottwoh[2]{
 \centering
 \includegraphics[width=0.45\hsize]{#1}
 \hfil
 \includegraphics[width=0.45\hsize]{#2}
}

\usepackage{graphicx}
\usepackage{amsmath}
\usepackage{amssymb}

\title [ Formation of spiral arms by swing amplification ]{ Global N-Body Simulation of Galactic Spiral Arms }
\author[S. Michikoshi and E. Kokubo]{
Shugo Michikoshi$^{1}$ \thanks{E-mail: michikos@kyoto-wu.ac.jp}
and Eiichiro Kokubo$^{2}$ \thanks{E-mail: kokubo@th.nao.ac.jp}
\\
$^{1}$ Department for the Study of Contemporary Society, Kyoto Women's University, Imakumano, Higashiyama, Kyoto, 605-8501, Japan \\
$^{2}$ Division of Theoretical Astronomy, National Astronomical Observatory of Japan, Osawa, Mitaka, Tokyo 181-8588, Japan
}

\date{Accepted XXX. Received YYY; in original form ZZZ}

\pubyear{2018}

\begin{document}
\label{firstpage}
\pagerange{\pageref{firstpage}--\pageref{lastpage}}
\maketitle

\begin{abstract}
The origin of galactic spiral arms is one of fundamental problems in astrophysics.
Based on the local analysis Toomre (1981) proposed the swing amplification mechanism in which the self-gravity forms spiral arms
 as leading waves of stars rotate to trailing ones due to galactic shear.
The structure of spiral arms is characterized by their number and pitch angle.
We perform global $N$-body simulations of spiral galaxies to investigate the dependence of the spiral structure on disk parameters and compare the simulation results with the swing amplification model.
We find that the spiral structure in the $N$-body simulations agrees well with that predicted by the swing amplification for the wide range of parameters.
The pitch angle decreases with increasing the shear rate and is independent of the disk mass fraction. 
The number of spiral arms decreases with both increasing the shear rate and the disk mass fraction. 
If the disk mass fraction is fixed, the pitch angle increases with the number of spiral arms.
\end{abstract}


\section{Introduction}

The formation mechanism of galactic spiral arms in disk galaxies is one
of important problems in galactic astronomy.
The spiral arms are excited by tidal interactions with nearby companion
galaxies \citep[e.g.,][]{Oh2008, Dobbs2010} and by the central stellar
bar \citep[e.g.,][]{Buta2005}. 
However, the spiral arms can also be excited and maintained without
external perturbations. 
One theory to explain the origin of spiral arms in disk galaxies is
swing amplification mechanism \citep{Goldreich1965, Julian1966,
Toomre1981}. 
During the rotation, a wave is amplified if Toomre's $Q$ is $1\mbox{\--}2$.
In $N$-body simulations of multi-arm spiral galaxies, it is observed
that the spiral arms are transient and recurrent
\citep[e.g.,][]{Sellwood1984, Sellwood2000, Baba2009, Fujii2011}. 
This feature can be understood by the swing amplification mechanism.

In a differentially rotating disk, if a perturber, such as a giant molecular cloud, exists,  
a stationary density structure around a perturber forms \citep{Julian1966}.
Even without a explicit perturber, the density pattern can be amplified. 
If the leading wave exists, it rotates to a trailing wave due to the shear.
If the self-gravity is sufficiently strong, the rotating wave is amplified during the rotation.
These processes are called swing amplification \citep{Goldreich1965, Julian1966, Toomre1981}.
The amplified density patterns may correspond to spiral arms observed in the galaxies.

In the swing amplification theory, the local and linear approximations were adopted.
First, the deviation of stellar orbits from the circular orbit on the disk midplane is assumed to be small compared to the orbital radius.
This is local approximation or epicycle approximation \citep{Binney2008}.
In addition, the deviation of various quantities, such as the surface density, from the unperturbed state is assumed to be small, that is, the deviation from the circular orbit is small compared to the wavelength.
Using this approximation, the hydrodynamic equation or Boltzmann equation in the local coordinate system is linearized \citep{Goldreich1965, Julian1966}.
In this respect, this is linear approximation.

In the linear theory of swing amplification, a perturber or a seed leading wave is necessary for the growth of the spiral arms.
\cite{Donghia2013} performed $N$-body simulations and examined the non-linear effect. 
Their simulations show that perturbers are not necessary once the spiral arms are developed. 
The spiral arm itself causes overdense and underdense regions that behave as perturbers and generate another spiral arm. 
This phenomenon cannot be explained only by the linear theory.
\cite{Kumamoto2016} clearly showed that the non-linear interaction between spiral arms forms overdense and underdense regions by the more controlled simulations.

However, we cannot still deny the importance of the linear theory of the swing amplification to explain the formation process of the spiral arm from a leading wave or a perturber caused by the non-linear interaction.
The linear theory of the swing amplification may explain some aspects of the basic physics of the spiral arm formation.
In addition, the short-scale spiral structures in Saturn's ring, so-called self-gravity wakes, are said to be formed by the swing amplification \citep{Salo1995, Michikoshi2015}.  
The recent $N$-body simulation suggests that self-gravity wakes exists even in a ring around a small body \citep{Michikoshi2017}.
This type of structure may be ubiquitous. 
Therefore, it is important to understand physical mechanism of swing amplification. 

In the series of our papers, we have investigated the swing
amplification mechanism using the local linear theory and the local
$N$-body simulations \citep{Michikoshi2014, Michikoshi2016, Michikoshi2016a}. 
The global $N$-body simulations of the spiral arms show that the pitch angle of
the spiral arms decreases with increasing the shear rate \citep{Grand2013}. 
This tendency is expected from the view of the swing amplification
mechanism \citep{Julian1966}. 
From the local $N$-body simulations and the local linear analyses, we
confirmed this trend and obtained the accurate pitch angle formula
\citep{Michikoshi2014} (hereafter referred to as Paper I). 
The proposed pitch angle formula is consistent with other global $N$-body simulations \citep{Grand2013, Baba2015, Fujii2018}.
The physical understanding of the dependence of the pitch angle on the
shear rate is given based on the phase synchronization of the epicycle
motion \citep{Michikoshi2016a} (hereafter referred to as Paper III). 

It is suggested that the number of spiral arms is inversely proportional to the disk mass
fraction \citep{Carlberg1985a}. 
\cite{Donghia2013} confirmed that the number of spiral arms is determined by the critical wavelength of the gravitational instability. 
It follows that the inverse relation between the disk mass and the number of spiral arm.
\cite{DOnghia2015} adopted more detailed model of disk and halo models and obtained the number of spiral arms formula, which depends on the distance from the galactic center.
Recently \cite{Fujii2018} performed global simulations that include
a live bulge and dark matter halo. Their results show that a larger
shear rate results in a smaller number of spirals. However, the
dependence of the number of spiral arms on the shear rate has not been
investigated quantitatively. 

In the previous works, a factor $X$, which is the azimuthal wavelength normalized by the critical wavelength, is assumed to be $1$--$2$. 
In general, $X$ depends on the shear rate \citep{Athanassoula1984}.
\cite{Michikoshi2016} (hereafter referred to as Paper II) obtained 
the detailed formula of $X$, the pitch angle, the amplification factor, and the number of spiral arms as a function of disk parameters.
It is suggested that $X$ increases and the number of spiral arms decreases with increasing the shear rate. 
This prediction has not yet been confirmed by global simulations.

So far the results of the local $N$-body simulations agree well with the local linear analysis of the swing amplification mechanism (Paper I, II).
However, in realistic spiral arms, the local approximation is not always valid.
Especially, for the grand-design spiral arms, the local approximation
would break down. 
Thus it is important to investigate the spiral structure by global
$N$-body simulations.    

In the present paper we extend local $N$-body simulations to global ones and  
systematically investigate the dependencies of the number and pitch angle of spiral arms on the disk parameters.
The outline of this paper is as follows. 
In Section \ref{sec:method}, we introduce the model and simulation method. 
In Section \ref{sec:res}, we give the results of the $N$-body simulations. 
In Section \ref{sec:dis}, we provide the intuitive explanation of the dependencies of the pitch angle and the number of spiral arms.
We summarize our findings in Section \ref{sec:summary}.

\section{Method} \label{sec:method}

\subsection{Model} \label{sec:model}

In the many previous works, Hernquist profile or NFW profile as the dark halo model is often adopted \citep{Hernquist1990, Navarro1997}. 
However, our aim is to examine the dependence on shear rate, disk mass fraction, and $Q$ and to understand the physical mechanism of swing amplification.
Thus, we introduce the somewhat artificial model for dark halo and disk to control these parameters directly, which is an straightforward extension of our local simulations presented in paper I, II.  
For the halo, we adopt the power-law model, which enables us to
control the shear rate directly 
\citep[e.g.,][]{Binney2008}. 
The similar dark halo model was also adopted in a recent controlled simulation \citep{Kumamoto2016}.
The density profile of the power-law density model follows
\begin{equation}
\rho_\mathrm{gh} = \rho_\mathrm{h0} \left(\frac{r}{r_0} \right)^{-\alpha},
\end{equation}
where $r$ is the distance from the center, $\alpha$ is the power-law
index of the density profile, $r_0$ is the typical scale length, and
$\rho_\mathrm{h0}$ is the density at $r_0$. 
Then, the corresponding orbital frequency is given as
\begin{equation}
\Omega_\mathrm{h} = \Omega_0 \left(\frac{r}{r_0} \right)^{-\alpha/2},
\label{eq:omega_h}
\end{equation}
where $\Omega_0$ is the orbital frequency at $r_0$.
The shear rate without the disk self-gravity is defined as 
\begin{equation}
\Gamma_\mathrm{h} =
 - \frac{\mathrm{d} \log \Omega_\mathrm{h}}{\mathrm{d} \log r}.
\end{equation}
If we neglect the disk self-gravity, the power-law density model gives
the uniform shear rate $\Gamma_\mathrm{h} = \alpha/2$.
Observationally disk galaxies have the shear rate $0.4 \lesssim \Gamma_\mathrm{h} \lesssim 1.5$ \citep{Seigar2005}.

The dark halo mass inside the sphere with radius $r$ is
$M_\mathrm{h} = r^3 \Omega_\mathrm{h}^2/G$.
From this, we can calculate the gravitational force from the dark halo at the point $\bm r$ with $|\bm r| = r$, where the origin is the center of the galaxy.
The gravitational acceleration by the halo is
$\bm {a}_\mathrm{h} = - \Omega_\mathrm{h}^2 \bm{r}$, which diverges for
$r \to 0$ when $\Gamma_\mathrm{h}>1/2$ ($\alpha>1$). 
In addition, the circular velocity diverges when
$\Gamma_\mathrm{h}>1$ ($\alpha>2$). 
For avoiding the divergence, in calculating the acceleration, we
introduce the softening parameter $\epsilon_\mathrm{h}$ as 
$\bm {a}_\mathrm{h} = -(GM_\mathrm{h}/(r^2+\epsilon_\mathrm{h}^2)^{3/2}) \bm{r}$.
We adopt $\epsilon_\mathrm{h} = 0.1 r_0$.
For $r \gg \epsilon_\mathrm{h}$, this modification does not affect the
result.  

We introduce the mass scale
\begin{equation}
M_\mathrm{h0} =  \frac{r_0^3\Omega_0^2}{G}.
\end{equation}
We normalize the length, time, and mass by $r_0$,
$\Omega_0^{-1}$, and $M_\mathrm{h0}$, respectively. 
In the following, the normalized quantities are denoted by a tilde on top.

The stellar disk surface density is given by an exponential model
\citep[e.g.,][]{Binney2008} 
\begin{equation}
\Sigma_\mathrm{d} = \Sigma_\mathrm{d0} e^{-\tilde r},
\label{eq:def_sigma_d}
\end{equation}
where $\Sigma_\mathrm{d0}$ is the surface density at $r_0$.
The mass inside the sphere with radius $r$ is
\begin{equation}
M_\mathrm{d}(r) =
M_\mathrm{d,\infty} \left(1 - \frac{1+\tilde r}{e^{\tilde r}}\right), 
\end{equation}
where $M_\mathrm{d,\infty}= 2 \pi \Sigma_\mathrm{d0} r_0^2$ is the
total mass of the stellar disk. 
In the following, we focus on the region with $1< \tilde r < 2$.
We define the disk mass fraction $f$ as
\begin{equation}
  f = \frac{M_\mathrm{d}(2r_0)}{ M_\mathrm{d}(2r_0) + M_\mathrm{h}(2r_0) }.
 \label{eq:def_f}
\end{equation}

It is often assumed that the ratio of the radial velocity dispersion to the surface density is constant \citep{Lewis1989, Hernquist1993}. 
Then, $Q$ depends on the distance from the galactic center.
However, the aim of this paper is to elucidate the dependence on the disk parameters.
Thus, we assume that the initial Toomre's $Q$, $Q_\mathrm{ini}$, of the disk is uniform. 
From this, we calculate the initial radial velocity dispersion $\sigma_r$.
The initial azimuthal velocity dispersion $\sigma_\theta$ is $\sigma_\theta/ \sigma_r = \kappa/2\Omega_\mathrm{h}$.
Here we assume that the initial vertical velocity dispersion $\sigma_z$ is given by an equilibrium ratio, which is  $\sigma_z/\sigma_r = 0.3 \kappa/\Omega_\mathrm{h}+0.2$, where $\kappa$ is the epicycle frequency for simplicity \citep[e.g.,][Paper I]{Ida1993}.
Using the epicycle approximation, we determine the initial velocity of
each particle with the random phase of the epicycle motion. 
Since the generated disk is not exactly in an equilibrium, the
artificial axisymmetric structure appears first.  
To remove this structure, we let the disk evolve for
$\tilde t/2 \pi = 15$ under the constraint of the rotational symmetry of the surface density by randomizing the azimuthal positions of particles 
\citep{McMillan2007, Fujii2011}.
Then we adopt it as the initial disk.

\begin{figure}
  \plotone{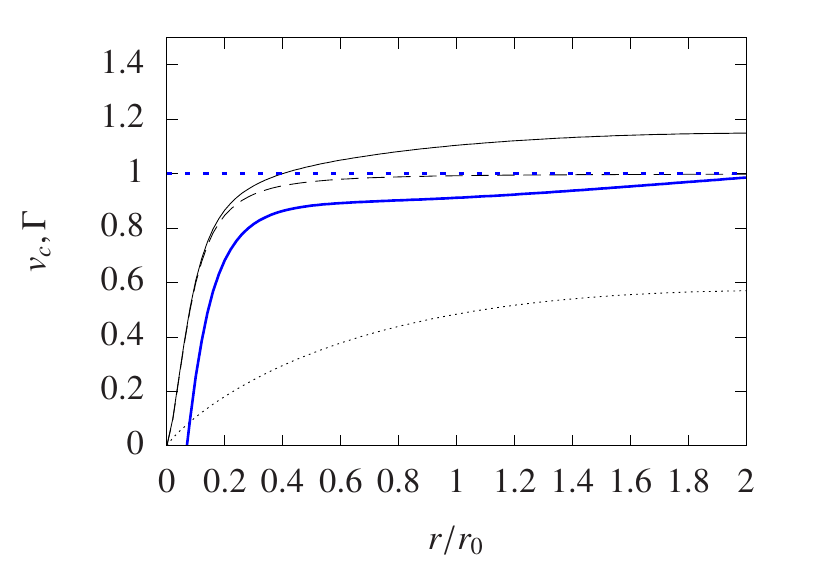}
  \caption{Circular velocity $v_\mathrm{c}$ and shear rate $\Gamma$ of model 7. The thin solid curve shows the circular velocity. The thin dashed and dotted curves show the circular velocity contributed by the halo and disk, respectively. The thick solid curve shows the shear rate calculated from the circular velocity. The thick dotted line shows $\Gamma_\mathrm{h}$.
   \label{fig:rotation_curve}
	}
\end{figure}

The number of stars is $N = 3 \times 10^6$. 
We introduce the softening length of the self-gravity between stars
$\epsilon$. 
We adopt $\tilde \epsilon = 0.01$, which is sufficiently small to
resolve the structures.  
In the models where the small scale structures appear ($f = 0.05$ and
$0.1$) we also perform simulations with $\tilde \epsilon = 0.005$ and
confirm that the following results do not change. 
In the following we vary $f$, $Q_\mathrm{ini}$, and $\Gamma_\mathrm{h}$.
The disk parameters are listed in Table~\ref{tbl:model}.
The models with $\Gamma_\mathrm{h} = 1$ (models 7, 9 and 13--30) correspond
to disks with flat rotation curve.
The circular velocity and shear rate of model 7 are shown in Fig.~\ref{fig:rotation_curve}.
The actual shear rate $\Gamma$ can deviate from $\Gamma_\mathrm{h}$
slightly due to the disk self-gravity and the velocity dispersion. 
From the averaged rotational velocity in the simulations, we calculate
$\Gamma$, which is summarized in Table~\ref{tbl:model}.
The difference between $\Gamma$ and $\Gamma_\mathrm{h}$ is small.
If $f$ is not so large, we can neglect the disk contribution to the
rotational velocity, which means $\Omega \simeq \Omega_\mathrm{h}$
where $\Omega$ is the orbital frequency. 
Then the actual shear rate is approximated by $\Gamma_\mathrm{h}$.

We use the simulation code based on FDPS, which is a general-purpose,
high-performance library for particle simulations
\citep{Iwasawa2016} with the Phantom-GRAPE module \citep{Tanikawa2012, Tanikawa2013}.
We adopt a leapfrog integrator with the fixed timestep $\Delta \tilde t = 2 \pi /1000$.

\begin{table*}
\centering
\caption{Disk Parameters and Results}
\label{tbl:model}
\begin{tabular}{cccccccc} 
  \hline
	Model & $f$ & $Q_\mathrm{ini}$ & $\Gamma_\mathrm{h}$ & $\Gamma$ & $Q_\mathrm{mean}$ & $\bar m$ & $\bar \theta$ \\
  \hline
1 & 0.20 & 1.2 & 0.4 & $0.48 \pm 0.03$ & $1.23 \pm 0.05$ & $10.2 \pm 2.5$ & $39.2 \pm 27.1$ \\
2 & 0.20 & 1.2 & 0.5 & $0.57 \pm 0.03$ & $1.27 \pm 0.07$ & $9.2 \pm 2.0$ & $36.1 \pm 13.2$ \\
3 & 0.20 & 1.2 & 0.6 & $0.65 \pm 0.03$ & $1.31 \pm 0.08$ & $8.1 \pm 1.0$ & $32.9 \pm 10.1$ \\
4 & 0.20 & 1.2 & 0.7 & $0.73 \pm 0.03$ & $1.37 \pm 0.10$ & $6.8 \pm 1.0$ & $29.9 \pm 9.2$ \\
5 & 0.20 & 1.2 & 0.8 & $0.81 \pm 0.03$ & $1.42 \pm 0.10$ & $6.1 \pm 1.0$ & $28.0 \pm 8.1$ \\
6 & 0.20 & 1.2 & 0.9 & $0.90 \pm 0.03$ & $1.46 \pm 0.11$ & $5.4 \pm 1.5$ & $25.7 \pm 7.9$ \\
7 & 0.20 & 1.2 & 1.0 & $0.98 \pm 0.03$ & $1.48 \pm 0.10$ & $4.8 \pm 1.0$ & $24.0 \pm 7.2$ \\
8 & 0.20 & 1.2 & 1.1 & $1.06 \pm 0.03$ & $1.48 \pm 0.11$ & $4.0 \pm 1.0$ & $20.7 \pm 8.2$ \\
9 & 0.20 & 1.2 & 1.2 & $1.14 \pm 0.02$ & $1.43 \pm 0.09$ & $3.6 \pm 1.0$ & $19.6 \pm 6.1$ \\
10 & 0.20 & 1.2 & 1.3 & $1.23 \pm 0.02$ & $1.42 \pm 0.10$ & $3.2 \pm 0.5$ & $16.9 \pm 6.0$ \\
11 & 0.20 & 1.2 & 1.4 & $1.31 \pm 0.01$ & $1.41 \pm 0.08$ & $2.5 \pm 0.5$ & $15.6 \pm 6.2$ \\
12 & 0.20 & 1.2 & 1.5 & $1.41 \pm 0.01$ & $1.48 \pm 0.06$ & $2.3 \pm 0.5$ & $13.3 \pm 4.5$ \\
13 & 0.40 & 1.2 & 0.4 & $0.53 \pm 0.07$ & $1.32 \pm 0.14$ & $4.6 \pm 1.0$ & $32.4 \pm 12.5$ \\
14 & 0.40 & 1.2 & 0.5 & $0.60 \pm 0.07$ & $1.36 \pm 0.13$ & $3.8 \pm 0.5$ & $31.0 \pm 8.3$ \\
15 & 0.40 & 1.2 & 0.6 & $0.67 \pm 0.06$ & $1.41 \pm 0.15$ & $4.4 \pm 1.0$ & $34.2 \pm 10.5$ \\
16 & 0.40 & 1.2 & 0.7 & $0.74 \pm 0.06$ & $1.44 \pm 0.14$ & $3.5 \pm 0.5$ & $25.6 \pm 15.8$ \\
17 & 0.40 & 1.2 & 0.8 & $0.80 \pm 0.06$ & $1.45 \pm 0.15$ & $3.5 \pm 0.5$ & $27.2 \pm 11.9$ \\
18 & 0.40 & 1.2 & 0.9 & $0.87 \pm 0.05$ & $1.44 \pm 0.13$ & $3.2 \pm 0.5$ & $25.4 \pm 9.6$ \\
19 & 0.40 & 1.2 & 1.0 & $0.94 \pm 0.04$ & $1.43 \pm 0.11$ & $2.8 \pm 0.5$ & $22.5 \pm 9.1$ \\
20 & 0.40 & 1.2 & 1.1 & $1.02 \pm 0.03$ & $1.38 \pm 0.07$ & $2.4 \pm 0.5$ & $20.4 \pm 10.6$ \\
21 & 0.40 & 1.2 & 1.2 & $1.10 \pm 0.03$ & $1.40 \pm 0.07$ & $2.4 \pm 0.5$ & $18.0 \pm 6.7$ \\
22 & 0.40 & 1.2 & 1.3 & $1.18 \pm 0.02$ & $1.40 \pm 0.06$ & $2.0 \pm 0.0$ & $15.5 \pm 6.2$ \\
23 & 0.40 & 1.2 & 1.4 & $1.27 \pm 0.03$ & $1.49 \pm 0.09$ & $2.0 \pm 0.0$ & $14.5 \pm 5.9$ \\
24 & 0.40 & 1.2 & 1.5 & $1.38 \pm 0.04$ & $1.75 \pm 0.14$ & $2.0 \pm 0.0$ & $13.8 \pm 10.3$ \\
25 & 0.05 & 1.2 & 1.0 & $0.99 \pm 0.01$ & $1.71 \pm 0.07$ & $14.5 \pm 3.0$ & $-14.3 \pm 35.4$ \\
26 & 0.10 & 1.2 & 1.0 & $0.99 \pm 0.02$ & $1.59 \pm 0.08$ & $8.0 \pm 1.0$ & $23.0 \pm 15.5$ \\
27 & 0.15 & 1.2 & 1.0 & $0.98 \pm 0.02$ & $1.52 \pm 0.10$ & $6.1 \pm 1.0$ & $24.5 \pm 7.2$ \\
28 & 0.20 & 1.2 & 1.0 & $0.98 \pm 0.03$ & $1.48 \pm 0.10$ & $4.8 \pm 1.0$ & $24.0 \pm 7.2$ \\
29 & 0.25 & 1.2 & 1.0 & $0.97 \pm 0.04$ & $1.45 \pm 0.11$ & $4.0 \pm 1.0$ & $22.6 \pm 8.3$ \\
30 & 0.30 & 1.2 & 1.0 & $0.96 \pm 0.04$ & $1.45 \pm 0.13$ & $3.5 \pm 0.5$ & $23.9 \pm 8.5$ \\
31 & 0.35 & 1.2 & 1.0 & $0.96 \pm 0.04$ & $1.42 \pm 0.12$ & $3.4 \pm 0.5$ & $23.5 \pm 7.6$ \\
32 & 0.40 & 1.2 & 1.0 & $0.94 \pm 0.04$ & $1.43 \pm 0.11$ & $2.8 \pm 0.5$ & $22.5 \pm 9.1$ \\
33 & 0.20 & 1.0 & 1.0 & $0.98 \pm 0.04$ & $1.78 \pm 0.12$ & $4.0 \pm 1.0$ & $25.4 \pm 9.2$ \\
34 & 0.20 & 1.1 & 1.0 & $0.98 \pm 0.03$ & $1.58 \pm 0.12$ & $4.6 \pm 0.5$ & $24.0 \pm 8.6$ \\
35 & 0.20 & 1.2 & 1.0 & $0.98 \pm 0.03$ & $1.48 \pm 0.10$ & $4.8 \pm 1.0$ & $24.0 \pm 7.2$ \\
36 & 0.20 & 1.3 & 1.0 & $0.98 \pm 0.02$ & $1.44 \pm 0.06$ & $5.0 \pm 1.0$ & $23.9 \pm 7.5$ \\
37 & 0.20 & 1.4 & 1.0 & $0.97 \pm 0.02$ & $1.47 \pm 0.03$ & $4.5 \pm 0.5$ & $22.6 \pm 9.1$ \\
38 & 0.20 & 1.5 & 1.0 & $0.97 \pm 0.02$ & $1.55 \pm 0.02$ & $4.4 \pm 0.5$ & $21.5 \pm 9.5$ \\
39 & 0.20 & 1.6 & 1.0 & $0.97 \pm 0.01$ & $1.63 \pm 0.03$ & $4.4 \pm 0.5$ & $21.2 \pm 16.7$ \\
40 & 0.20 & 1.7 & 1.0 & $0.97 \pm 0.01$ & $1.73 \pm 0.03$ & $5.2 \pm 1.5$ & $19.9 \pm 19.3$ \\
41 & 0.20 & 1.8 & 1.0 & $0.97 \pm 0.01$ & $1.82 \pm 0.04$ & $5.1 \pm 2.5$ & $16.9 \pm 19.4$ \\
42 & 0.20 & 1.9 & 1.0 & $0.97 \pm 0.01$ & $1.92 \pm 0.04$ & $8.7 \pm 12.0$ & $13.8 \pm 34.3$ \\
  \hline
\end{tabular}
\end{table*}

\subsection{Analysis of Spiral Arms}

In order to analyze the spiral arms quantitatively, we calculate their number
and pitch angle using the Fourier coefficients.
As stated in Section \ref{sec:model}, we focus on the disk region of
$1.0 < \tilde r < 2.0$ where the softening effect of the acceleration from the halo is negligible. 
We divide the region into 10 annuli with width $\Delta \tilde r = 0.1$.
In each annulus, we calculate the surface density
$\Sigma(\tilde r,\phi, \tilde t)$, where $\phi$ is the azimuthal angle.
The Fourier coefficient at radius $\tilde r$ for $m \ge 1$ is defined by
\begin{equation}
A_m(\tilde r, \tilde t) = \frac{1}{\pi} \int_{0}^{2\pi} \frac{\Sigma(\tilde r,\phi, \tilde t)}{\bar \Sigma(\tilde r, \tilde t)} \exp ( - i m \phi) \mathrm{d} \phi,
\end{equation}
where $\bar \Sigma(\tilde r, \tilde t)$ is the azimuthally averaged
surface density,  
\begin{equation}
\bar \Sigma(\tilde r, \tilde t) = \frac{1}{2\pi} \int_0^{2\pi}
\Sigma(\tilde r, \phi, \tilde t) \mathrm{d}\phi.
\end{equation}
For $m=0$, the Fourier coefficient is $A_0 = 1$.
Conversely, we can reconstruct the surface density from
$A_m(\tilde r,\tilde t)$ 
\begin{equation}
\frac{\Sigma(\tilde r,\phi, \tilde t)}{\bar \Sigma(\tilde r, \tilde t)} =
\sum_{m=0}^\infty |A_m(\tilde r, \tilde t)| \cos( m \phi +  \phi_m), 
\end{equation}
where $\phi_m = \tan^{-1} \left( \Im( A_m(\tilde r, \tilde t)) / \Re(
A_m(\tilde r, \tilde t) ) \right)$ is the shape function. 

\begin{figure*}
 \includegraphics[width=0.9\hsize]{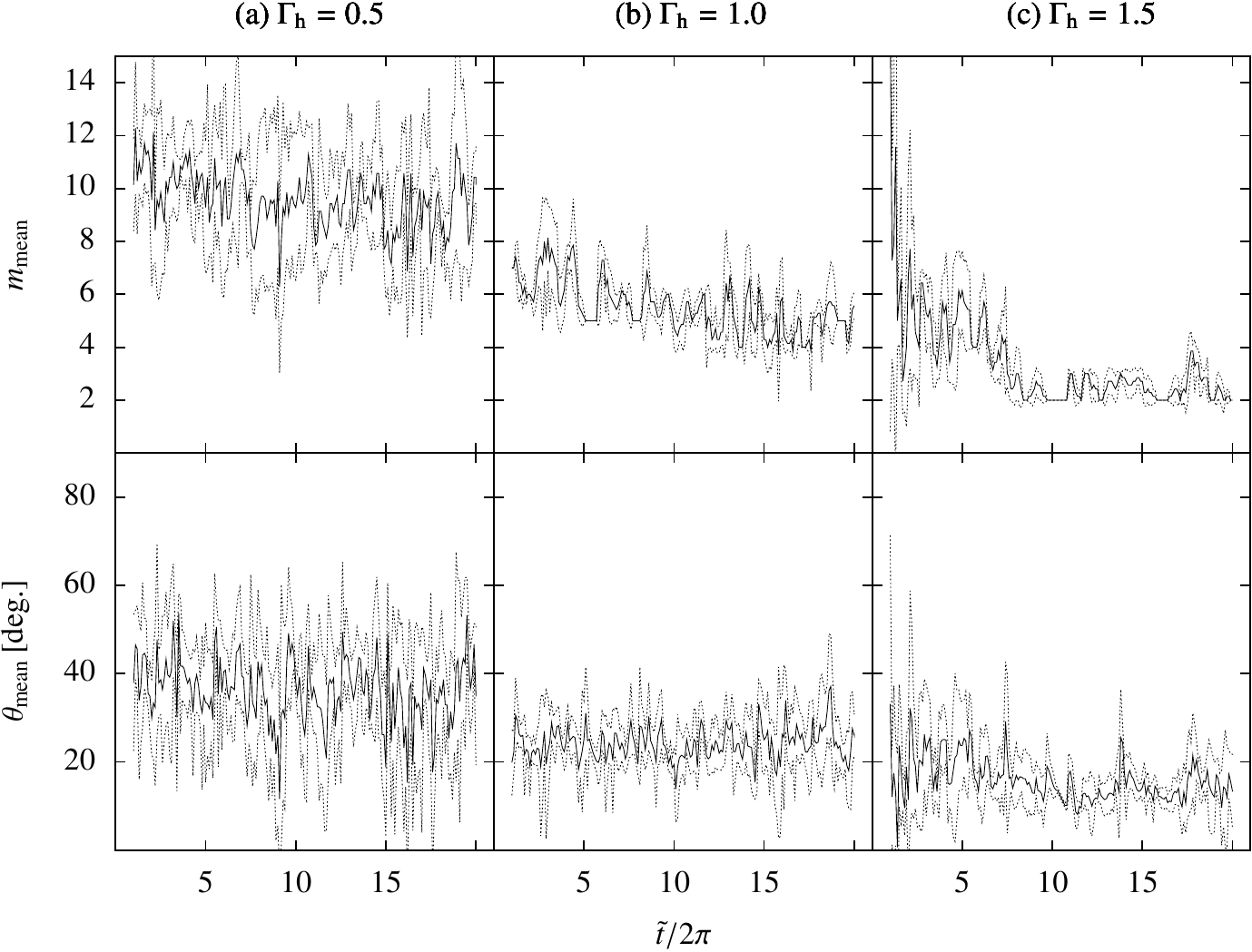}
 \caption{Time evolution of the averaged number of spiral arms $m_\mathrm{mean}(\tilde t)$ and the averaged pitch angle $\theta_\mathrm{mean}(\tilde t)$ for (a) $\Gamma_\mathrm{h}=0.5$ (model 2), (b) $\Gamma_\mathrm{h}=1.0$ (model 7), and (c) $\Gamma_\mathrm{h}=1.5$ (model 12).
	The solid curves represent $m_\mathrm{mean}(\tilde t)$ and $\theta_\mathrm{mean}(\tilde t)$.
	The dotted curves represent the quartile deviation.
	\label{fig:timevo_m}
	}
\end{figure*}

We estimate the number of spiral arms using Fourier coefficient.
First we find $m_\mathrm{max}(\tilde r, \tilde t)$ that has the maximum
amplitude $|A_m(\tilde r, \tilde t)|$ of the annulus with radius $\tilde r$ at $\tilde t$.
Since the number of spiral arms varies with the radius \citep{Fujii2011}, $m_\mathrm{max}(\tilde r, \tilde t)$ has the relatively large dispersion.
Thus, we calculate the mean value of $m_\mathrm{max} (\tilde r, \tilde t)$
at each $\tilde t$, $ m_\mathrm{mean}(\tilde t) = \langle m_\mathrm{max} (\tilde r, \tilde t) \rangle_{\tilde r}$, where $\langle X \rangle_Y$ denotes the average of $X$ with respect to the variable $Y$.
In calculating average, we adopt the interquartile mean for avoiding the influence from outliers. 
Fig.~\ref{fig:timevo_m} shows the time evolution of $m_\mathrm{mean}(\tilde t)$. 
For $\Gamma_\mathrm{h} = 0.5$, $m_\mathrm{mean}(\tilde t)$ is almost constant
throughout the simulation, which is about $\simeq 9$.
On the other hand, for $\Gamma_\mathrm{h} = 1.0$ and $1.5$, $m_\mathrm{mean}(\tilde t)$
decreases with time until $\tilde t/ 2\pi \simeq 10$ and then becomes
almost constant.
Thus we calculate the number of spiral arms by the time average over $10 < \tilde t / 2\pi< 20$, $\bar m = \langle m_\mathrm{max}(\tilde r, \tilde t) \rangle_{\tilde r, \tilde t}$.

The pitch angle $\theta_m$ for mode $m$ is defined by the shape function
\citep[e.g.,][]{Binney2008}, 
\begin{equation}
\cot \theta_m(\tilde r, \tilde t) = \frac{\tilde r}{m} \frac{\mathrm{d} \phi_m}{\mathrm{d} \tilde r} \simeq \frac{\tilde r}{2 m \Delta \tilde r} (\phi_m(\tilde r + \Delta \tilde r, \tilde t)-\phi_m(\tilde r - \Delta \tilde r, \tilde t)) .
\end{equation}
We calculate $\theta_\mathrm{max}(\tilde r, \tilde t)$
of the dominant mode with $m_\mathrm{max}(\tilde r, \tilde t)$. 
We calculate $\theta_\mathrm{mean}(\tilde t) = \langle \theta_\mathrm{max}(\tilde r, \tilde t) \rangle_{\tilde r}$ by averaging $\theta_\mathrm{max}(\tilde r, \tilde t)$ over $\tilde r$.
Fig.~\ref{fig:timevo_m} shows the time evolution of $ \theta_\mathrm{mean}(\tilde t)$.
Although the pitch angle  has the large dispersion, there seems no clear trend. 
Finally we obtain the time average of the pitch angle 
over $10 < \tilde t / 2\pi < 20$, $\bar \theta = \langle \theta_\mathrm{max}(\tilde r,\tilde t)\rangle_{\tilde r, \tilde t}$.

\begin{figure}
 \begin{center}
	  \includegraphics[width=\columnwidth]{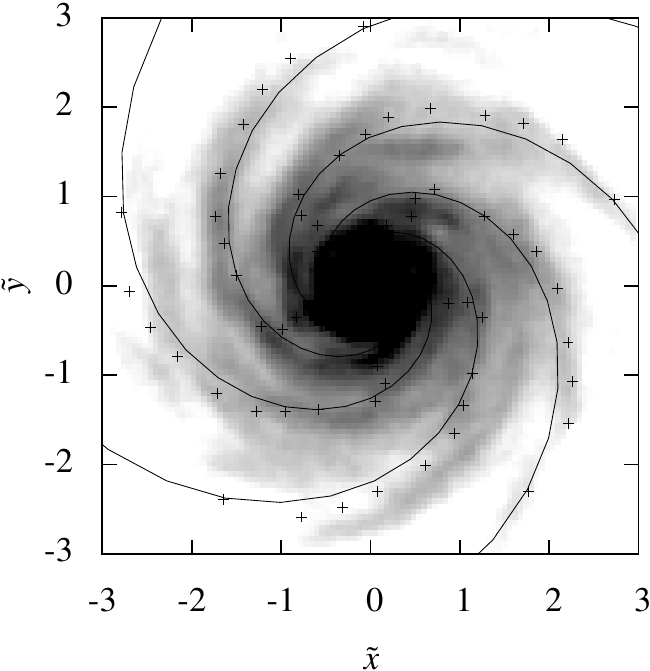}
 \end{center}
	\caption{The density snapshot and the estimated spiral arms for $\Gamma_\mathrm{h}=1.0$, $f=0.2$, and $Q_\mathrm{ini}=1.2$ (model 7) at $\tilde t/ 2\pi=15.0$.
	The plus points denote the position of spiral arm estimated from the shape function $\phi_m$ with $m=5$.
	The solid curves show the logarithmic spirals with $m=5$ and $\theta=24 ^\circ$.
	  \label{fig:spiralarm} 
	}
\end{figure}

Fig.~\ref{fig:spiralarm} demonstrates the estimated spiral arms for the 
surface density distribution at $\tilde t / 2\pi = 15$ for model 7 where
$\Gamma_\mathrm{h}=1.0$, $f=0.2$, and $Q_\mathrm{ini}=1.2$.
Their estimated number and pitch angle are $\bar m = 4.8$ and $\bar \theta = 24.0 ^\circ$,
which are consistent with the numerical results.
Assuming $m = 5$ and $\theta = 24.0 ^\circ$, we draw the logarithmic spiral.
The $j$-th logarithmic spiral ($j=0, 1, 2, \cdots, m-1$) is given by
\begin{equation}
\phi =  - \frac{\phi_m(r_0)}{m} - \frac{1}{\tan \theta} \log \left(\frac{r}{r_0} \right) + \frac{2 \pi j}{m},
\end{equation}
where $\phi_m(r_0)$ is the phase at radius $r_0$.
We find that the logarithmic spiral arms with the estimated spiral
parameters agree with the simulation.
This indicates that the dependence of the pitch angle on the
galactocentric distance is weak.  

\section{Results} \label{sec:res}

\subsection{Spiral Arm Structures} \label{sec:armst}

\begin{figure*}
	\begin{minipage}{0.33\hsize}
	  \plotone{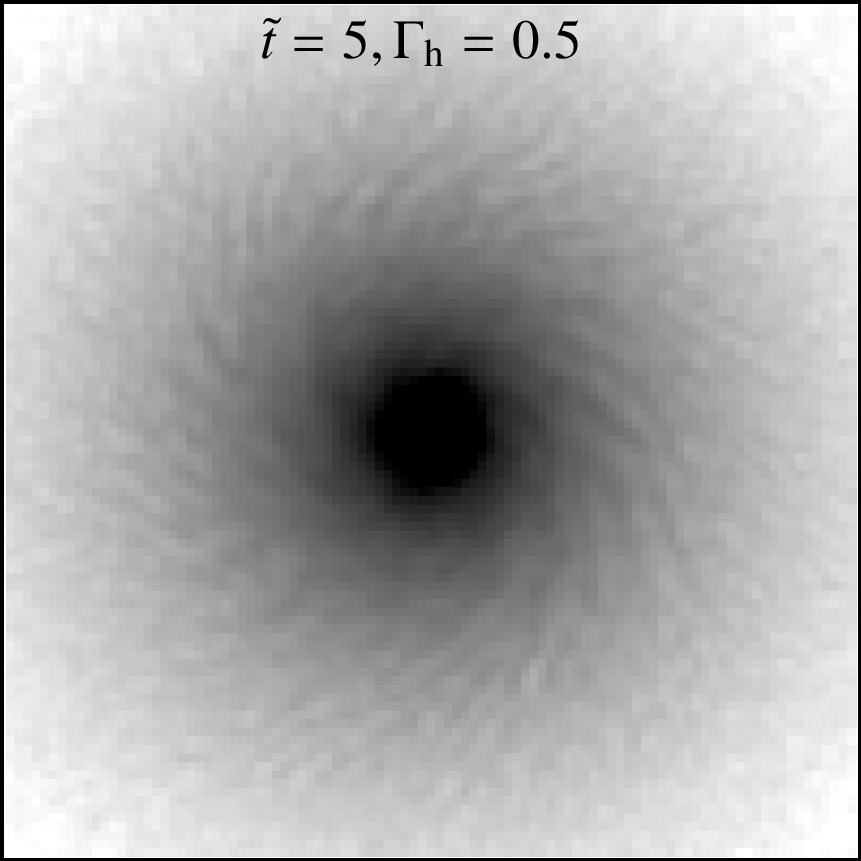}
	\end{minipage}
	\begin{minipage}{0.33\hsize}
	  \plotone{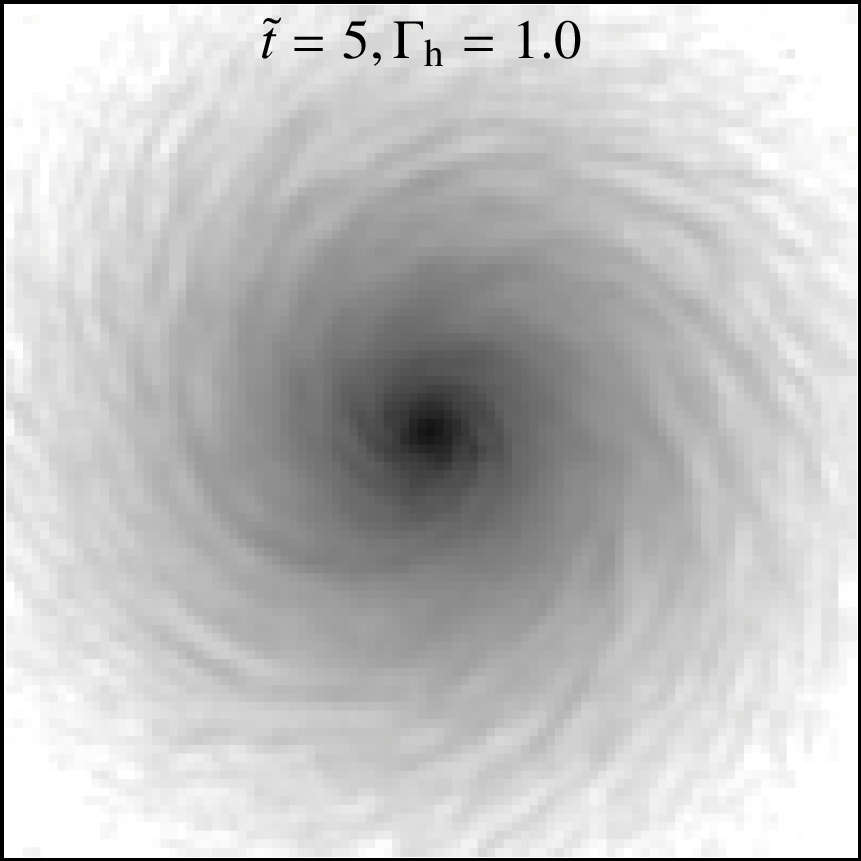}
	\end{minipage}
	\begin{minipage}{0.33\hsize}
	  \plotone{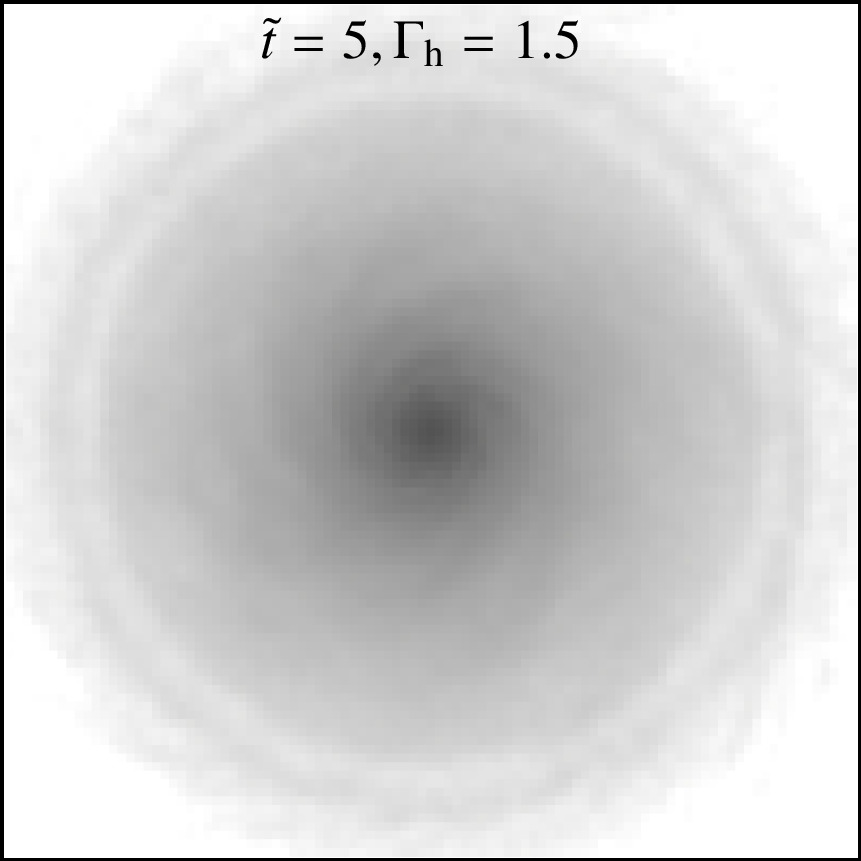}
	\end{minipage}

	\begin{minipage}{0.33\hsize}
	  \plotone{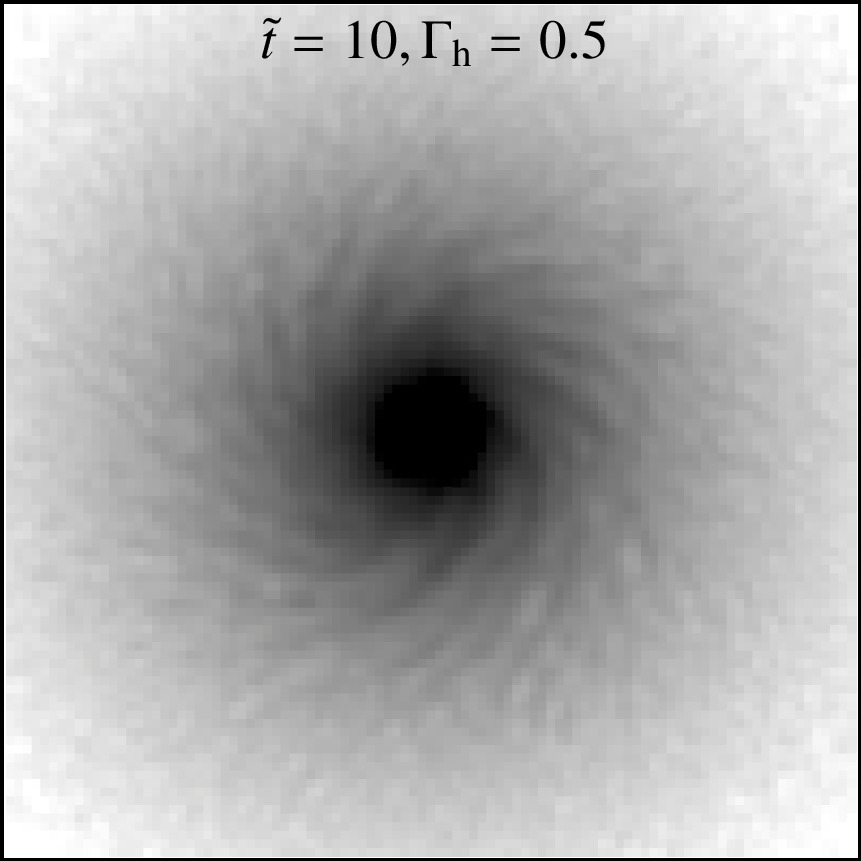}
	\end{minipage}
	\begin{minipage}{0.33\hsize}
	  \plotone{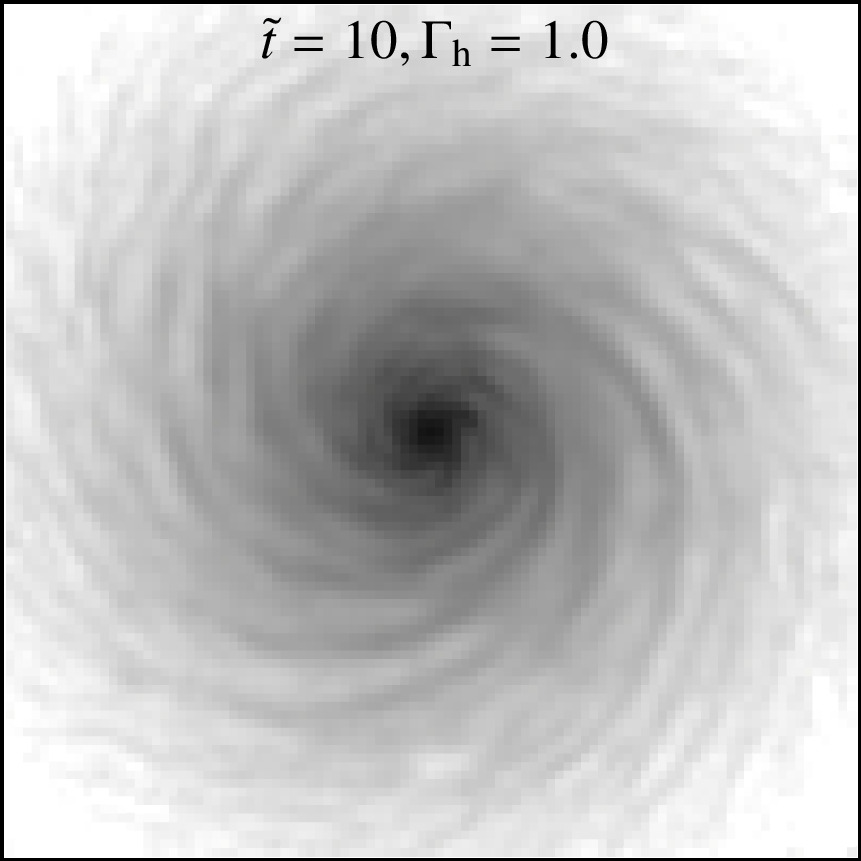}
	\end{minipage}
	\begin{minipage}{0.33\hsize}
	  \plotone{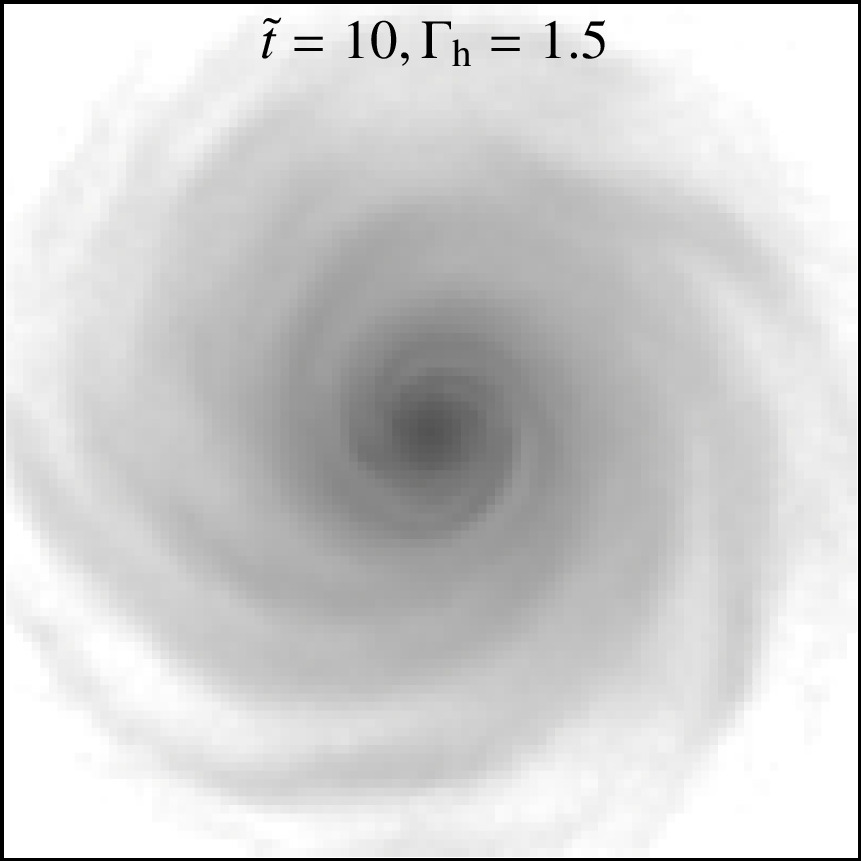}
	\end{minipage}

	\begin{minipage}{0.33\hsize}
	  \plotone{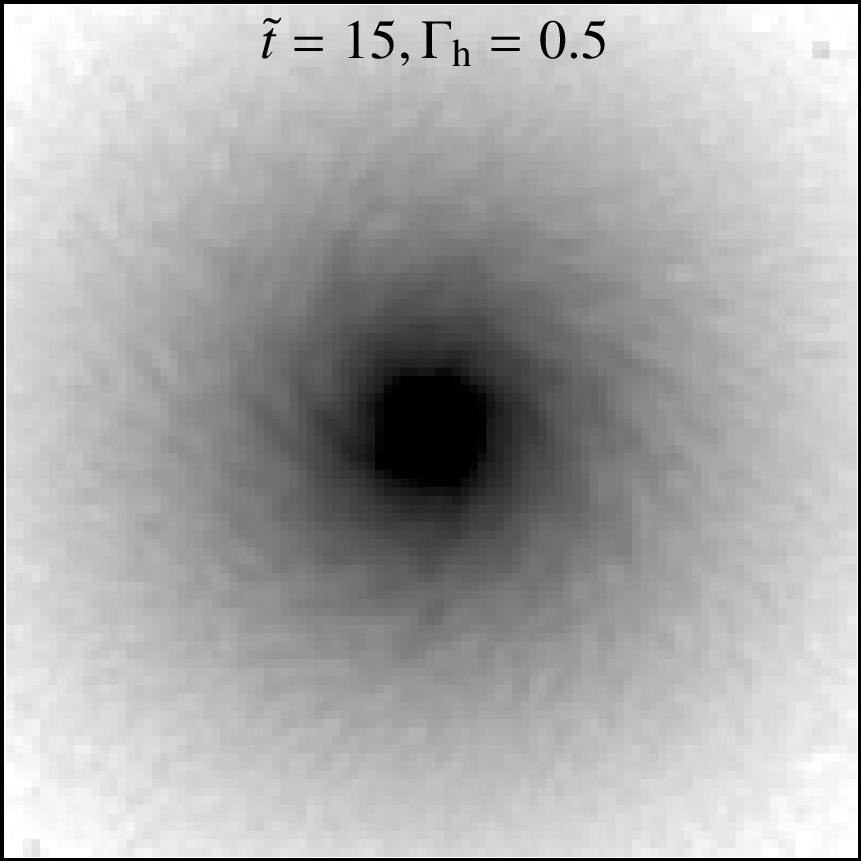}
	\end{minipage}
	\begin{minipage}{0.33\hsize}
	  \plotone{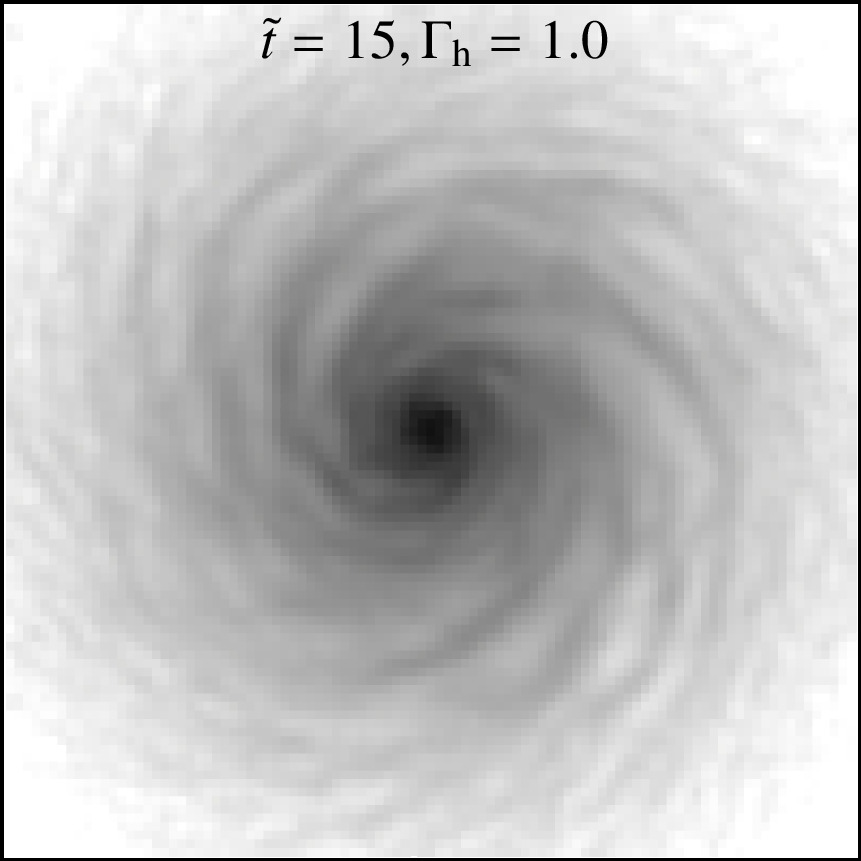}
	\end{minipage}
	\begin{minipage}{0.33\hsize}
	  \plotone{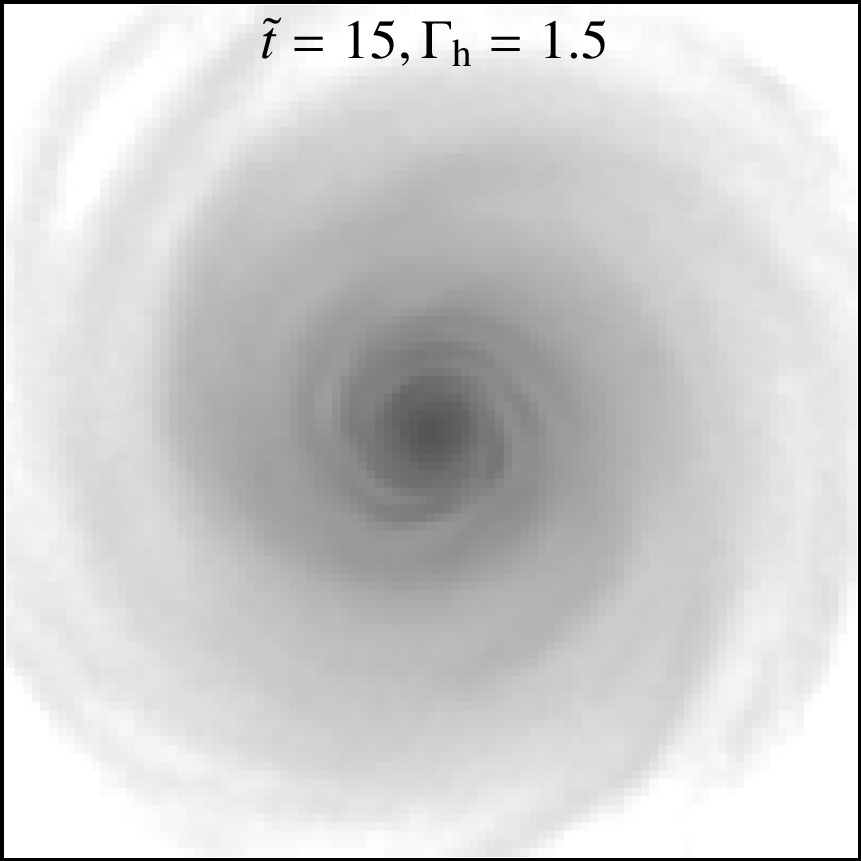}
	\end{minipage}

\caption{
  Snapshots of the surface density for $\Gamma_\mathrm{h}=0.5$ (model 2), $1.5$
(model 7), and $1.5$ (model 12) at $\tilde t/2\pi = 5.0, 10.0$ and $15.0$. 
The surface density is shown in logarithmic scale in the region of
$-5 \le \tilde x \le 5$ and $-5 \le \tilde y \le 5$.
\label{fig:spiral_structure}
}
\end{figure*}

Fig.~\ref{fig:spiral_structure} shows the spiral arm structures for
$\Gamma_\mathrm{h}=0.5$ (model 2), $\Gamma_\mathrm{h}=1.0$ (model 7), and $\Gamma_\mathrm{h}=1.5$ (model 12)
where $f = 0.2$.
The spiral arms are transient and recurrent, that is, the spiral arms
are formed and destructed continuously.
We find that the overall spiral arm structures, such as the pitch angle
and the number, barely change with time for $\tilde t/2 \pi > 10$.

For small $\Gamma_\mathrm{h}$, the length scale of the spiral arms is short and
their number is large, while for large $\Gamma_\mathrm{h}$, the length scale is
long and their number is small. 
Namely, the number of spiral arms decreases with increasing $\Gamma_\mathrm{h}$.
For larger $\Gamma_\mathrm{h}$, the spiral arms are wound more tightly, in other
words, the pitch angle is smaller.

\begin{figure}
	\plotone{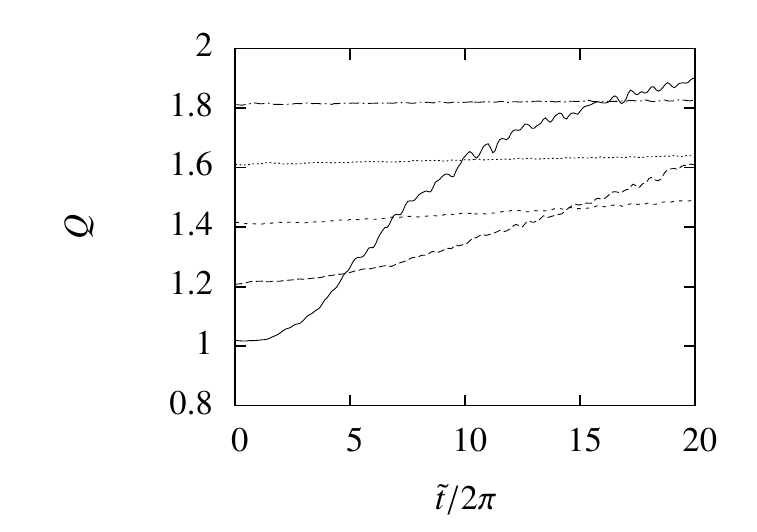}
\caption{
Time evolution of $Q$ for $Q_\mathrm{ini} = 1.0, 1.2, 1.4, 1.6,$
and $1.8$  (models 33, 35, 37, 39, 41).
\label{fig:q_evolution}
}
\end{figure}

Fig.~\ref{fig:q_evolution} shows the evolution of $Q$ for
$Q_\mathrm{ini}=1.0, 1.2, 1.4, 1.6,$ and $1.8$ (models 33, 35, 37, 39, 41).
In a way similar to the calculation of $\bar \theta$ and $\bar m$, we calculate the average of $Q$ .
For smaller $Q$, $Q$ increases more rapidly.
This is consistent with the previous $N$-body simulations
\citep[][Paper I]{Fujii2011}. 
In the case of smaller $Q$, the amplification factor is larger and the
spiral arms are denser \citep{Toomre1981}. 
Since the stars are scattered by the denser spiral arms more strongly, 
$Q$ increases more rapidly.
The time-averaged $Q$ over $\tilde t / 2 \pi = 10\mbox{--}20$ for each
model is summarized in Table~\ref{tbl:model}.

\subsection{Comparison with Swing Amplification Theory}

We examine the dependencies of $\bar \theta$ and $\bar m$ on $\Gamma$, $f$, and $Q$.
In Papers I and II, based on the swing amplification, $\bar m$ and $\bar \theta$  are
estimated as 
\begin{equation}
\bar m =  0.922 C \frac{(2-\Gamma)^2}{f Q},
\label{eq:numarm0}
\end{equation}
\begin{equation}
\tan \bar \theta =
\frac{1}{2 \pi} \left(1+ \frac{2.095}{Q^{5.3}} \right)^{-1}\frac{\kappa}{A},
\label{eq:pitch}
\end{equation}
where $A$ is the Oort constant. 
In deriving equation (\ref{eq:numarm0}) we assumed
that the orbital frequency is given by $\Omega^2 \simeq \pi C G \Sigma_\mathrm{d}/ f r$ with a fudge factor $C$ of order unity.
Note that in reality $C$ depends on the disk and halo models (see Appendix \ref{sec:estfac}).
For the range of $Q = 1.5\mbox{--}1.8$, equation (\ref{eq:pitch}) is
reduced to  
\begin{equation}
\tan \bar \theta \simeq
\frac{1}{7} \frac{\kappa}{A} = \frac{2 }{7} \frac{\sqrt{4-2 \Gamma}}{\Gamma}.
\label{eq:pitch2}
\end{equation}

\subsubsection{Dependence on $\Gamma$}

\begin{figure*}
 \includegraphics[width=0.9\hsize]{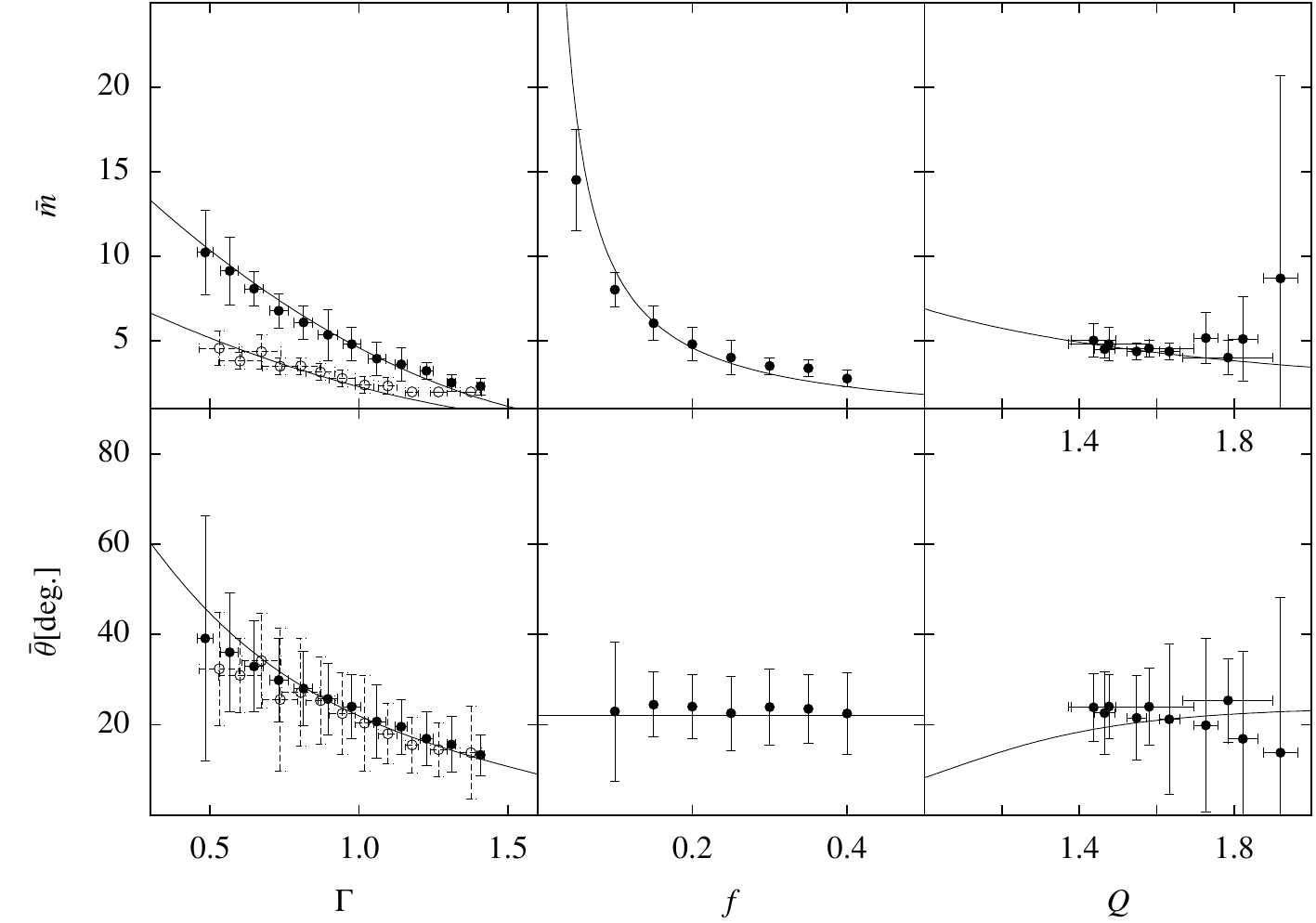}
\caption{
  Number of spiral arms $\bar m$ and the pitch angle and $\bar \theta$ against $\Gamma$, $f$, and $Q$.
  In the top panels (number of spiral arms $\bar m$), the solid curves show equation (\ref{eq:numarm0}) with $C = 1.5$.
  In the bottom panels (pitch angle $\theta$), the solid curves show equation (\ref{eq:pitch2}). 
  In the left panels ($\Gamma$--dependence), the filled and open circles correspond to the models of $f=0.2$ (models 1--12) and $f=0.4$ (models 25--32), respectively.
The error bars represent quartile deviation.
\label{fig:depedence}
}
\end{figure*}

Fig.~\ref{fig:depedence} shows the results of the $N$-body simulations.
From Table~\ref{tbl:model}, we adopt $Q \simeq 1.5$ for equations (\ref{eq:numarm0}). 

The number of spiral arms $\bar m$ decreases with increasing $\Gamma$, which
is consistent with equation (\ref{eq:numarm0}). 
We find that equation (\ref{eq:numarm0}) with $C=1.5$ agrees well with the numerical results.

The pitch angle $\bar \theta$ also decreases with increasing $\Gamma$.
We find that equation (\ref{eq:pitch2}) agrees well with the numerical results.

\cite{Fujii2018} performed the $N$-body simulations with a more realistic
galactic model that includes a live bulge and dark matter halo. They compared
the pitch angle with equation (\ref{eq:pitch2}) and concluded that
equation (\ref{eq:pitch2}) agrees with the $N$-body simulations. 
Therefore, it is suggested that equation (\ref{eq:pitch2}) is applicable under
the general galactic models. In addition, they reported that a larger shear
rate results in a smaller number of spirals. Our simulation results
and equation (\ref{eq:numarm0}) are also consistent with their results.

\subsubsection{Dependence on $f$}

Equation (\ref{eq:numarm0}) with $C = 1.5$ shows that $\bar m$ decreases with increasing $f$,
which agrees with the $N$-body simulations. 
This result is consistent with the previous works \citep{DOnghia2015, Fujii2018}.

Equation (\ref{eq:pitch2}) indicates that the pitch angle does not
depend on $f$, which is confirmed by the $N$-body simulations. 
If we adopt $\Gamma = 1.0$ and $Q_\mathrm{ini} = 1.2$, the mean pitch angle
is $22 ^\circ \mbox{--} 25 ^\circ$ and is independent of $f$ 
except for $f=0.05$.
In the case of the model with $f=0.05$, the structure is too small and faint to calculate accurately the pitch angle.

Note that in the models adopted here, $\Gamma$ and $f$ are varied
independently. Thus $\bar \theta$ is completely independent of $f$.
Usually, a galaxy model with high $f$ tends to have high $\Gamma$, where
$\bar \theta$ depends on $f$ through $\Gamma$.

\subsubsection{Dependence on $Q$}

Equations (\ref{eq:numarm0}) and (\ref{eq:pitch2}) show that $\bar m$
decreases and $\bar \theta$ increases with increasing $Q$, but their dependencies are weak. 
The numerical results are consistent with the swing amplification.

\subsubsection{$\theta$--$m$ Relation}

\begin{figure}
\plotone{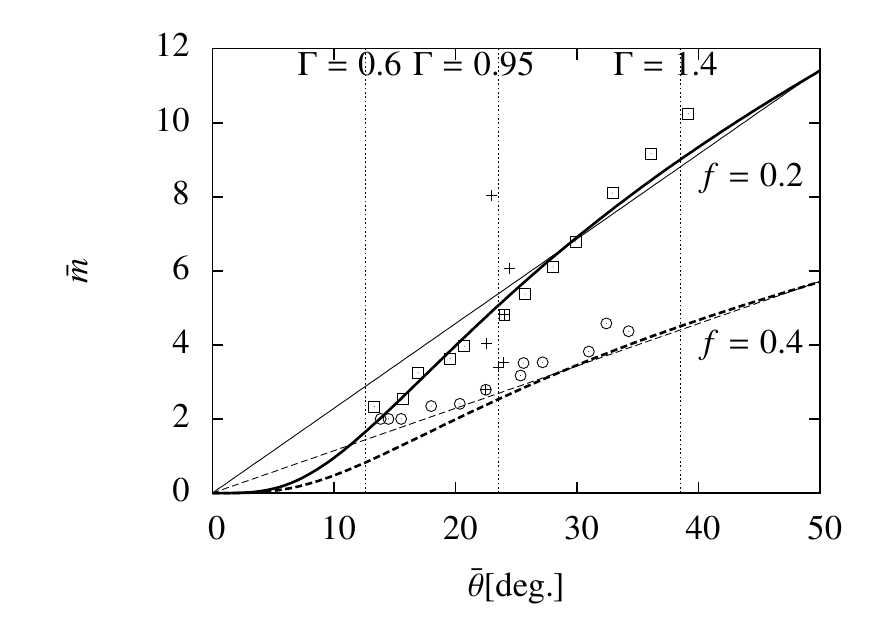}
\caption{
Pitch angle $\bar \theta$ against the number of the spiral arms $\bar m$ for the
models of $f=0.2$ (models 1--12, squares), $f=0.4$ (models 13--24, circles), and $\Gamma_\mathrm{h} =1.0$ (models 25--32, pluses).
The solid and dashed thick curves show the estimates given by equation
(\ref{eq:mtrelation}) with $f = 0.2$ and $f=0.4$, respectively.
The factor $C = 1.5$ and $Q=1.5$ are adopted.
The thin lines show the approximated estimate given by equation (\ref{eq:mtrelation2}). 
The vertical dotted lines show equation (\ref{eq:pitch2}) with $\Gamma=0.6$, $0.95$, and $1.4$.
\label{fig:m_vs_th}
}
\end{figure}

We investigate the relation between the pitch angle and the number of
spiral arms. 
The swing amplification mechanism gives its relation as (Paper II)
\begin{equation}
\bar m \simeq
\frac{0.230 C}{Q f}
\left( \frac{14 \tan \bar \theta}{1+\sqrt{1+49 \tan^2 \bar \theta}} \right)^{4}.
\label{eq:mtrelation}
\end{equation}
For $\bar \theta>30^\circ$, this can be approximated by
\begin{equation}
  \bar m \simeq
6.1 C \left(\frac{\bar \theta}{40^\circ}\right)
\left(\frac{f}{0.2}\right)^{-1} \left(\frac{Q}{1.5} \right)^{-1}.
\label{eq:mtrelation2}
\end{equation}
Fig.~\ref{fig:m_vs_th} shows the results of $N$-body simulations with $f = 0.2$ (models 1--12), $f=0.4$ (models 13--24) and $\Gamma_\mathrm{h} =1.0$ (models 25--32).
In equations (\ref{eq:mtrelation}) and (\ref{eq:mtrelation2}), we adopt $Q=1.5$ and $C=1.5$.
The swing amplification mechanism generally agrees with the $N$-body simulations.
Thus we predict that the pitch angle increases with the number
of spiral arms if the spiral arms are formed by the swing
amplification mechanism.

From the observational data, the positive correlation between the pitch angle and the number of spiral arms of unbarred multi-arm spiral galaxies has been reported \citep{Hart2017}.
This correlation is consistent with equation (\ref{eq:mtrelation}).
In order to confirm this relation in the observational studies more quantitatively, 
it is necessary to analyze $\theta$--$m$ relation together with $f$--dependence. 

\section{Discussion} \label{sec:dis}

We present the intuitive derivation of the pitch angle and number of
spiral arms.  
Except for the numerical coefficient, the pitch angle formula can be
obtained from the phase synchronization argument \citep{Michikoshi2016a}. 
We briefly summarize its derivation.
We consider a single leading wave in a rotating frame.
Due to the shear, the wave rotates from leading to trailing.
When the wave changes from leading to trailing, 
the stabilizing effect of Coriolis force is reduced.
Thus, the particles are pulled towards the wave center by the
self-gravity and their epicycle phases are synchronized. 
Then the wave amplitude becomes the maximum after the half of an epicycle period \citep{Michikoshi2014}. 
The pitch angle evolves with time as $\tan \theta = 1/(2A t)$ where $t$
is the elapsed time from $\theta = 90^\circ$. 
Substituting $t = \pi/\kappa$ we obtain the pitch angle as
$\tan \theta \simeq \kappa/2\pi A$.
This result is consistent with that of the local simulations and
the swing amplification $\tan \theta \simeq \kappa/7A$ \citep{Michikoshi2014}.

The azimuthal wavelength $\lambda_y$ is given by the pitch angle $\theta$ and the radial
wavelength $\lambda_x$,  
\begin{equation}
\lambda_y = \frac{\lambda_x}{\tan \theta}.
\end{equation}
The radial wavelength is often assumed to be
$\lambda_x = \lambda_\mathrm{cr}$, where $\lambda_\mathrm{cr}$ is the
critical wavelength of the gravitational instability for the
axisymmetric modes \citep{Toomre1964}.
Though this relation is not obvious for non-axisymmetric modes,
the local linear analyses of the swing amplification and the local $N$-body simulations confirm
$\lambda_x \simeq \lambda_\mathrm{cr}$ for $\kappa/\Omega<1.6$
\citep{Michikoshi2016}.  
Thus, adopting this relation we obtain
\begin{equation}
\lambda_y =
\frac{\lambda_\mathrm{cr}}{\tan \theta} \simeq
\frac{28 \pi^2 G \Sigma_\mathrm{d} A}{\kappa^3}. 
\end{equation}
Using the azimuthal wavelength, we calculate the number of spiral arms as
\begin{equation}
m =
\frac{2 \pi r}{\lambda_y} \simeq \frac{\kappa^3 r}{14 \pi G \Sigma_\mathrm{d} A} \sim
\frac{\kappa^3}{14 f A \Omega^2},
\end{equation}
where we used the approximation
$\Omega^2 \simeq \pi G \Sigma_\mathrm{d}/r f$ (Paper II, Appendix  \ref{sec:estfac}).
For $1.0 < \kappa/\Omega <1.5$, we numerically find that $A \kappa / \Omega^2 $ is almost constant between 0.65 and 0.77.
Thus in this parameter range, we can approximate $A \kappa / \Omega^2$ as a constant $0.71$. 
This approximation can recover the previous result, which is 
\begin{equation}
m \sim 0.1 \frac{\kappa^4}{f \Omega^4}.
\end{equation}
This expression agrees with that obtained by the swing amplification 
(equation (\ref{eq:numarm0})) except for the dependence on $Q$. 

In the above argument we assumed that
$\lambda_x \simeq \lambda_\mathrm{cr}$. 
On the other hand, $\lambda_y$ increases with $\Gamma$ and decreases
with increasing $\theta$ since $\theta$ decreases with increasing
$\Gamma$.  
This indicates that $m$ is larger for larger $\theta$, which is
consistent with equation (\ref{eq:mtrelation}).

\section{Summary} \label{sec:summary}

We have performed the global $N$-body simulations of disk galaxies in
order to compare the spiral structure with those by the swing
amplification theory. 
The mean pitch angle and number of spiral arms were calculated in the
disks with various shear rates and mass fractions.
We confirmed that the dependencies of the spiral structure on
disk parameters agree with those in the swing amplification
theory.  
The pitch angle decreases with increasing the shear rate and is
independent of the disk mass fraction. 
The number of spiral arms decreases with both increasing the shear rate
and the disk mass fraction. 
It follows that the pitch angle tends to increases with the number of spiral arms if the disk mas fraction is fixed.

From the swing amplification mechanism only we cannot understand the
 overall process of the spiral arm formation.
The $N$-body simulations show that the spiral arms are transient and
 recurrent, that is, the spiral arms are formed and destructed continuously.
Two questions remain unsolved in this process.
One is the origin of seed leading waves.
In the realistic galaxies, the swing amplification mechanism requires relatively strong leading waves.
The $N$-body simulations show that the overdense or underdense regions forms due to the nonlinear interaction between spiral arms \citep{Donghia2013, Kumamoto2016}. 
However, its physical mechanism is still unclear. 
We have to understand the generation mechanism such leading waves.
The other is the fate of the amplified spiral arms.
The $N$-body simulations show that the amplified arms are finally
 destructed.
The destruction mechanism has not yet been understood completely.
\cite{Baba2013} pointed out that the stars in spiral arms escape and the
spiral arms damp due to the non-linear wave-particle interaction. It
is also suggested that the nonlinear wave-wave interaction generates the
leading arms from the swing-amplified arms \citep{Fuchs2005}. 
The wave-wave interaction may also contribute to damping of 
the spiral arms. In addition, the gas component of the disk neglected
in the present study, may potentially affect the dynamics of spiral
arms \citep[e.g.,][]{Bottema2003}. Further study on this effect is necessary.

In our $N$-body simulations, we adopt a artificial galactic model and generate the initial condition by simple manner to control the key disk parameters.
Our results suggest that a fudge factor in equation (\ref{eq:numarm0}) depends on the galactic model.
Thus, it is necessary to determine this factor for more realistic galactic model. 
In addition, the initial condition of our model is not exactly in an equilibrium. 
Thus, we adopt the randomizing-azimuthal method for avoiding unnatural structures \citep{McMillan2007, Fujii2011}.
It would be better to adopt the more sophisticated method for generating initial conditions \citep{Hernquist1993, Kuijken1995, McMillan2007, Miki2018}.
In the future work, we will validate the swing amplification theory base on more realistic model.

Numerical computations were carried out on Cray XC30 at Center for Computational Astrophysics, National Astronomical Observatory of Japan.

\section*{Acknowledgements}

Numerical computations were carried out on ATERUI (Cray XC30) at the Center for Computational Astrophysics, National Astronomical Observatory of Japan.

%

\bibliographystyle{mnras}

\appendix
\section{Estimation of Factor $C$} \label{sec:estfac}
We describe the approximation that we employed in deriving the number of spiral arms in Paper II.
Considering the azimuthal wavelength $\lambda_y$, the number
 of spiral arms is written as
\begin{equation}
  m = \frac{2 \pi r}{\lambda_y} = \frac{1}{2.17 \cdot 2\pi Q} \frac{r \Omega^2}{G\Sigma_\mathrm{d}} \left(\frac{\kappa}{\Omega} \right)^4,
  \label{eq:mest}
\end{equation}
where we used $\lambda_y = 2.17 Q (\Omega / \kappa)^2 \lambda_\mathrm{cr}$ (Paper II).
We assume that the orbital frequency is 
\begin{equation}
  \Omega^2 \simeq \frac{G M_\mathrm{tot}}{r^3},
\label{eq:omega}
\end{equation}
where $M_\mathrm{tot}$ is the total mass inside the sphere with radius $r$.
In addition, we assume that the disk mass inside the sphere with radius $r$ is roughly given by $\pi r^2 \Sigma_{\mathrm{d}}$.
Thus, $M_\mathrm{tot}$ is given as
\begin{equation}
  M_\mathrm{tot} \simeq \frac{ \pi r^2 \Sigma_{\mathrm{d}}}{f}.
\label{eq:mtot}
\end{equation}
Substituting equation (\ref{eq:mtot}) into equation (\ref{eq:omega}), 
we obtain 
\begin{equation}
  \Omega^2 = \frac{\pi C G \Sigma_{\mathrm{d}}}{f r}, 
\label{eq:omega2}
\end{equation}
where we introduce a fudge factor $C$.
Substituting equation (\ref{eq:omega2}) into equation (\ref{eq:mest}), we obtain the number of spiral arms with a factor $C$ (equation (\ref{eq:numarm0})).
We expect that $C$ is an order of unity, though its value depends on the disk and halo models.
In what follows, we estimate $C$ assuming an exponential disk and a power-law halo models.

\begin{figure*}
  \plottwoh{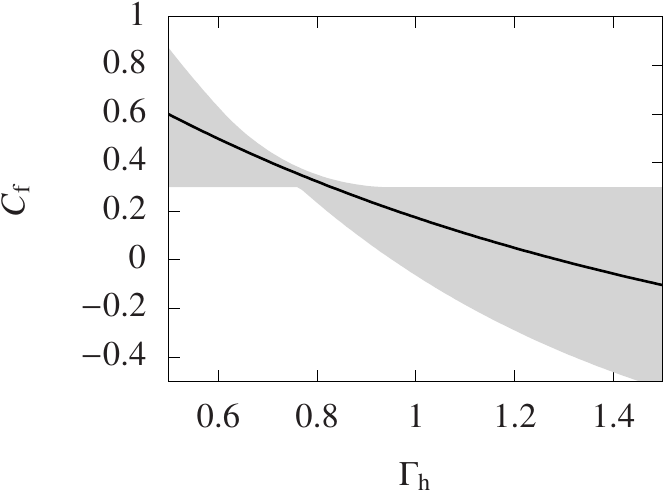}{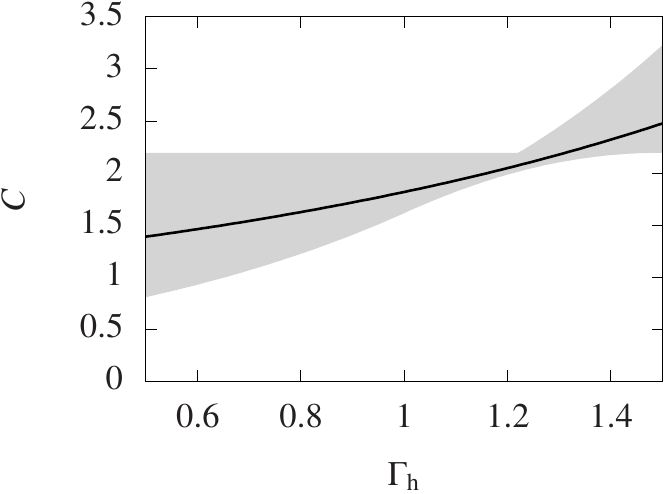}
  \caption{Factors $C_\mathrm{f}$ and $C$ as a function of $\Gamma_\mathrm{h}$. 
    The solid curves show the values averaged over $\tilde r = 1 \mbox{--} 2$.
The shaded regions show the minimum and maximum values for $\tilde r = 1 \mbox{--} 2$.
In calculating $C$, we assumed $C_f f = 0$.
    \label{fig:factor_c_ch}
	}
\end{figure*}

The orbital frequency is separated into two components, 
\begin{equation}
  \Omega^2 = \Omega_\mathrm{h}^2  + \Omega_\mathrm{d}^2,
  \label{eq:omega_sep}
\end{equation}
 where $\Omega_\mathrm{d}$ is the contribution by the disk given by
\begin{equation}
  \Omega_\mathrm{d}^2 = \frac{\pi G \Sigma_\mathrm{d0}}{r_0} (I_0(\tilde r/2) K_0(\tilde r/2) - I_1(\tilde r/2) K_1(\tilde r/2)),   
  \label{eq:omega_disk}
\end{equation}
 where $I_\nu$ and $K_\nu$ are the modified Bessel functions of the first
 and second kinds and $\nu$ is an order \citep{Freeman1970}.
 Substituting equations (\ref{eq:omega_h}) and (\ref{eq:omega_disk}) into equation (\ref{eq:omega_sep}), we obtain  
\begin{equation}
  C = \frac{f r \Omega^2}{\pi G \Sigma_\mathrm{d}} = \frac{e^{\tilde r}}{2} \left(1 - \frac{3}{e^2}\right) \left(\frac{\tilde r}{2} \right)^{1-2 \Gamma_\mathrm{h}} \left(1 + C_f f\right),
\end{equation}
 where 
\begin{equation}
C_\mathrm{f} = \frac{4(I_0(\tilde r/2) K_0(\tilde r/2) - I_1(\tilde r/2) K_1(\tilde r/2))}{ 1 - 3/e^2} \left(\frac{\tilde r}{2} \right)^{2\Gamma_\mathrm{h}}-1.
\end{equation}
As shown in Fig.~\ref{fig:factor_c_ch}, $C_\mathrm{f}$ depends on $\Gamma_\mathrm{h}$ and $\tilde r$, which ranges from $-0.5$ to $0.8$.
 The averaged value over $1 < \tilde r < 2$ and $0.5 < \Gamma_\mathrm{h} < 1.5$ is about $0.20$.
 Thus, if $f$ is not large, we can use the approximation $C_f f \simeq 0$.
The factor $C$ with $C_f f \simeq 0$ is shown in Fig.~\ref{fig:factor_c_ch}.
We obtain that $C$ averaged over for $1 < \tilde r < 2$ and $0.5 < \Gamma_\mathrm{h} < 1.5$ is $1.85$.
Therefore, $C \simeq 1.85$ is a good approximation.

\bsp	
\label{lastpage}
\end{document}